\documentclass[sigconf]{acmart}

\usepackage{amsmath}
\usepackage{enumitem}
\usepackage{bbm}
\usepackage{pifont}
\usepackage{array}
\usepackage{graphicx}
\usepackage{multirow}
\usepackage{booktabs}
\usepackage[table]{xcolor}

\definecolor{baselinebg}{HTML}{D9F0F1}
\definecolor{oursbg}{HTML}{ECE6F7} 
\definecolor{improvpos}{HTML}{1A7F4B}
\definecolor{improvneg}{HTML}{C0392B}

\AtBeginDocument{%
  }

\setcopyright{acmlicensed}
\copyrightyear{2018}
\acmYear{2018}
\acmDOI{XXXXXXX.XXXXXXX}

\acmConference[Conference acronym 'XX]{Make sure to enter the correct
  conference title from your rights confirmation email}{June 03--05,
  2018}{Woodstock, NY}

\acmISBN{978-1-4503-XXXX-X/2018/06}

\begin{document}

\title{Learning to Rank Explanations for Recommendation with Language Models}

\author{Ben Kabongo}
\email{ben.kabongo@sorbonne-universite.fr}
\affiliation{%
  \institution{Sorbonne University, CNRS, ISIR}
  \city{Paris}
  \country{France}
}

\author{Vincent Guigue}
\email{vincent.guigue@agroparistech.fr}
\affiliation{%
  \institution{AgroParisTech, UMR MIA Paris-Saclay}
  \city{Palaiseau}
  \country{France}}

\renewcommand{\shortauthors}{Ben Kabongo \& Vincent Guigue}

\begin{abstract}
Recommender systems are ubiquitous on the Web, making
understandable explanations essential for users.
Although large language models (LLMs) increasingly
generate natural-language justifications, evaluating
their factual grounding and personalized relevance
remains challenging. Statement-level ranking offers
an alternative by selecting concise, review-derived
explanations, enabling fine-grained and reproducible
evaluation using established ranking metrics.
However, existing statement ranking methods largely rely on collaborative
representations, while semantic enhancements often fail
to outperform simple user-history popularity baselines.
Moreover, directly adapting language models for
personalized statement ranking remains largely unexplored.
In this work, we introduce \textsc{LMStaR} (Language Models for
Statement Ranking), a fully text-based framework that
formulates personalized explanation ranking as dense
retrieval. A query encoder represents the user's
interaction history and target-item metadata, while
a statement encoder maps candidate
explanations into a shared embedding space.
Both encoders are jointly adapted through multi-positive
contrastive learning to learn personalized relevance.
Experiments on Amazon Beauty, Amazon Toys, and Yelp
show that \textsc{LMStaR} achieves the highest NDCG@1
across all three domains, outperforming \textsc{CLER}
by 1.06--8.25 points (+11.8--65.9\%).
It also leads top-five ranking on both Amazon datasets,
although \textsc{CLER} remains stronger on Yelp.
Task-specific adaptation improves all five evaluated
backbones and transfers across domains without
target-domain fine-tuning. Counterfactual analyses
further reveal the importance of user-specific histories. These findings highlight the
potential of task-adapted language models for
personalized explanation ranking.
Code, data, and supplementary materials are available at
\url{https://github.com/BenKabongo25/language-models-statement-ranking}.
\end{abstract}



\maketitle

\section{Introduction}
\label{sec:introduction}

Recommender systems play a central role on Web platforms,
helping users discover products, services, and content.
As increasingly sophisticated models underpin these
recommendations, providing understandable justifications
is important both for users assessing suggested items
and for designers seeking to diagnose system
behavior~\cite{zhang2020explainable}.
Textual explanations are particularly appealing because
they convey justifications in natural, human-understandable
language. Existing approaches have evolved from predefined
templates~\cite{li2020generate} to neural
generation~\cite{li2021personalized} and, more recently,
LLM-based generation~\cite{li2023personalized,ma2024xrec}.
However, these generative methods may produce generic
or unsupported rationales~\cite{kabongo2025factual}.
Evaluating their factual grounding with semantic similarity
metrics~\cite{zhang2019bertscore} or LLM-based
judges~\cite{sheng2025analyzing}, as well as assessing
the relevance of individual explanatory factors, remains
challenging.
Such concerns extend beyond recommendation, motivating
evidence-grounded approaches in question
answering~\cite{stolfo2024groundedness} and retrieval-based
tool selection in LLM agents~\cite{qin2025meta}.
In explainable recommendation, a complementary paradigm
is \emph{statement-level ranking}: rather than generating
free-form explanations, approaches such as
\textsc{EXTRA}~\cite{li2021extra} and
\textsc{StaR}~\cite{kabongo2026rank} rank concise,
review-derived explanatory statements and return the
top-$k$ for each user--item interaction.
This formulation enables individual explanatory aspects
to be prioritized and supports reproducible evaluation
using established ranking metrics~\cite{jarvelin2002cumulated}.

Yet ranking explanations does not inherently solve their
personalization. Under \emph{item-level} evaluation, candidates
are restricted to the target item's statements, making
the challenge to identify which aspects best match a
particular user. Prior work shows that simple user-history
popularity signals remain competitive with sophisticated
rankers in this setting~\cite{kabongo2026rank}.
Existing approaches range from neighborhood-based
collaborative filtering and factorization methods~\cite{li2021extra} to
\textsc{BPER} and its variants~\cite{li2023relationship}. To leverage textual semantics,
\textsc{BPER+}~\cite{li2023relationship} incorporates BERT representations, while
\textsc{SE-BPER} and \textsc{SE-BPER+} exploit contextualized
statement embeddings to initialize user, item, and
statement factors~\cite{xu2022learning}.
These approaches retain an identifier-based collaborative
formulation, predominantly optimized with Bayesian
Personalized Ranking~\cite{rendle2012bpr}.
More recently, \textsc{CLER}~\cite{escarda2026contrastive}
introduces contrastive learning through an NT-BXent
objective, substantially improving collaborative ranking
without explicitly modeling textual semantics.
While effective, these approaches leave the direct,
task-specific alignment of textual user histories and
candidate explanations largely unexplored.

Meanwhile, neural information retrieval has progressed
from dual-encoder architectures such as
{Sentence-BERT}~\cite{reimers2019sentence} and
{DPR}~\cite{karpukhin2020dense} to adapting large
language models as text encoders and rankers.
Approaches such as {RepLLaMA}~\cite{ma2024fine}, {LLM2Vec}\cite{behnamghader2024llm2vec},
and {Qwen3-Embedding}~\cite{zhang2025qwen3} demonstrate the effectiveness
of task-adapted language-model representations for
retrieval.
By independently encoding queries and documents,
bi-encoders enable similarity-based ranking with
precomputable candidate representations, while contrastive
learning aligns embeddings with task-specific relevance.
Related advances in recommendation, including
{UniSRec}~\cite{hou2022towards} and {RecFormer}~\cite{li2023text}, exploit textual
item representations and interaction sequences.
These advances motivate exploring language models
as explanation rankers within a fully text-based
framework for personalized statement ranking.

In this work, we introduce \textbf{L}anguage \textbf{M}odels for \textbf{Sta}tement
\textbf{R}anking (\textsc{LMStaR}), a fully text-based framework that
formulates personalized explanation ranking as a dense
retrieval problem. To our knowledge, \textsc{LMStaR} is
the first approach to jointly adapt language-model
encoders for personalized statement ranking.
Rather than learning representations tied to user and
item identifiers, our approach directly models the
textual context of each user--item interaction.
Specifically, a query encoder represents the user's
interaction history and target-item metadata, while
a separate statement encoder independently maps
candidate explanations into a shared embedding space.
Both encoders are jointly optimized using a multi-positive
contrastive objective, contrasting relevant statements
against item-level negatives to learn personalized relevance.

We construct three statement-ranking datasets from user
reviews in Amazon Beauty, Amazon Toys~\cite{ni2019justifying},
and Yelp, using a refined \textsc{StaR}
pipeline~\cite{kabongo2026rank} that emphasizes statements
describing matches between user preferences and item
characteristics, and strengthens reranker-based verification
of statement cluster merges.
We compare against popularity-based~\cite{kabongo2026rank},
neighborhood-based~\cite{li2021extra},
factorization~\cite{li2021extra,li2023relationship},
semantic-enhanced~\cite{li2023relationship,xu2022learning},
contrastive collaborative~\cite{escarda2026contrastive},
and zero-shot retrieval~\cite{zhang2025qwen3} baselines.
We further investigate five pretrained backbones,
LoRA~\cite{hu2022lora} versus full fine-tuning,
and cross-domain transfer without target-domain adaptation.
To characterize the role of user context, we conduct
controlled analyses of history quantity and selection,
performance across history-length groups, and
counterfactual query perturbations.

Our results show that \textsc{LMStaR} achieves the
highest NDCG@1 across all three domains, outperforming
\textsc{CLER} by 3.50 (+31.2\%), 8.25 (+65.9\%),
and 1.06 (+11.8\%) points on Beauty, Toys, and Yelp,
respectively. At $k=5$, it also improves NDCG by
2.66 (+13.4\%) and 7.79 (+34.2\%) points on Beauty
and Toys, while trailing \textsc{CLER} by 1.10
($-6.7$\%) points on Yelp.
Beyond overall effectiveness, task-specific adaptation
improves all evaluated backbones across domains,
with smaller adapted models outperforming larger
zero-shot alternatives. LoRA further exceeds full
fine-tuning by 0.84--2.66 NDCG@5 points across the
six tested configurations.
These gains also transfer across domains: without
target-domain fine-tuning, adapted models outperform
zero-shot retrieval and conventional ranking baselines
in all six transfer settings.
Finally, history analyses reveal substantial gains from
the first few recent interactions, while replacing the
user's history with another user's reduces NDCG@5 points across the three domains.
In contrast, perturbing target-item metadata has a
comparatively limited effect, highlighting the importance
of user-specific context under item-level ranking.

Our main contributions are summarized as follows:
\begin{itemize}[leftmargin=*,nosep]
    \item We introduce \textsc{LMStaR}, a fully text-based
    dual-encoder framework that formulates personalized
    statement ranking as dense retrieval and jointly
    adapts language-model representations through
    multi-positive contrastive learning.

    \item We demonstrate the effectiveness and robustness of
    \textsc{LMStaR} across three recommendation domains,
    examining ranking performance against established
    baselines, five pretrained backbones, fine-tuning
    strategies, and cross-domain transferability.

    \item We systematically investigate the effects of user
    history quantity, selection, and length, together
    with counterfactual query perturbations, revealing
    the importance of user-specific interaction histories
    and the comparatively limited influence of target-item
    metadata under item-level ranking.
\end{itemize}

\section{Related Work}
Our work connects text-based explainable recommendation
with neural retrieval.

\subsubsection*{Text-based Explainable Recommendation}
Textual explanations have evolved from templates
~\cite{li2020generate} to RNNs~\cite{dong2017learning,li2017neural},
Transformers~\cite{li2021personalized}, and LLM-based
generation~\cite{li2023personalized,ma2024xrec,li2025g}.
However, lexical metrics emphasize surface overlap
~\cite{papineni2002bleu,lin2004rouge}, semantic metrics
do not ensure factuality~\cite{zhang2019bertscore,
kabongo2025factual}, and LLM-based judgments remain
difficult to reproduce~\cite{fu2023gptscore,sheng2025analyzing}.
Statement ranking~\cite{li2021extra,kabongo2026rank}
instead selects the top-$k$ explanatory statements,
enabling fine-grained evaluation with ranking metrics
~\cite{jarvelin2002cumulated}, provided statements are
explanatory, atomic, and semantically unique.

\subsubsection*{Statement Ranking Methods}
Existing methods include popularity baselines
(\textsc{UserPop}, \textsc{ItemPop}, \textsc{GlobalPop})
~\cite{kabongo2026rank}, neighborhood models
(\textsc{RUCF}, \textsc{RICF}), and factorization
(\textsc{CD}, \textsc{PITF})~\cite{li2021extra}.
\textsc{BPER} learns user--statement and item--statement
scores~\cite{li2023relationship}.
Its semantic variants include \textsc{BPER+}, combining
latent factors with BERT embeddings~\cite{li2023relationship},
and \textsc{SE-BPER}/\textsc{SE-BPER+}, initializing
factors from contextualized statements, with task-specific
fine-tuning in the latter~\cite{xu2022learning}.
Unlike these BPR-based methods~\cite{rendle2012bpr},
\textsc{CLER}~\cite{escarda2026contrastive} employs
contrastive NT-BXent learning while remaining ID-based.
Ranking may be \emph{global-level}, over all statements,
or \emph{item-level}, restricted to the target item
~\cite{kabongo2026rank,li2021extra}.
The latter isolates personalization more directly:
on \textsc{StaR}, \textsc{UserPop} matches or exceeds
several item-level rankers~\cite{kabongo2026rank}.

\subsubsection*{Statement Ranking Datasets}
\textsc{EXTRA}~\cite{li2021extra} extracts review sentences
and clusters near-duplicates using lexical similarity
and locality-sensitive hashing~\cite{anand2011mining},
potentially missing semantic paraphrases.
\textsc{StaR}~\cite{kabongo2026rank} instead combines
LLM-based extraction with semantic retrieval, reranking,
and graph-based clustering.
We refine its extraction prompts toward user--item matches
and strengthen verification of cluster merges.

\subsubsection*{Neural Retrieval Models}
Dual encoders such as {DSSM}~\cite{huang2013learning},
{Sentence-BERT}~\cite{reimers2019sentence},
and {DPR}~\cite{karpukhin2020dense}
learn shared query--document embedding spaces.
Subsequent advances include hard-negative contrastive
learning~\cite{xiong2020approximate}, weak supervision
~\cite{wang2022text}, and instruction-conditioned
embeddings~\cite{su2023one}.
{RepLLaMA}~\cite{ma2024fine},
{LLM2Vec}~\cite{behnamghader2024llm2vec},
and, more recently, {Qwen3-Embedding}~\cite{zhang2025qwen3}
extend retrieval to language-model representations.
In recommendation, {UniSRec}~\cite{hou2022towards},
{RecFormer}~\cite{li2023text}, and LLM-based
two-tower models~\cite{xu2026case} leverage textual
interaction contexts for item retrieval.
\textsc{LMStaR} instead adapts these representations
to personalized explanation ranking.

\begin{figure*}
\centering
\includegraphics[width=1\linewidth]{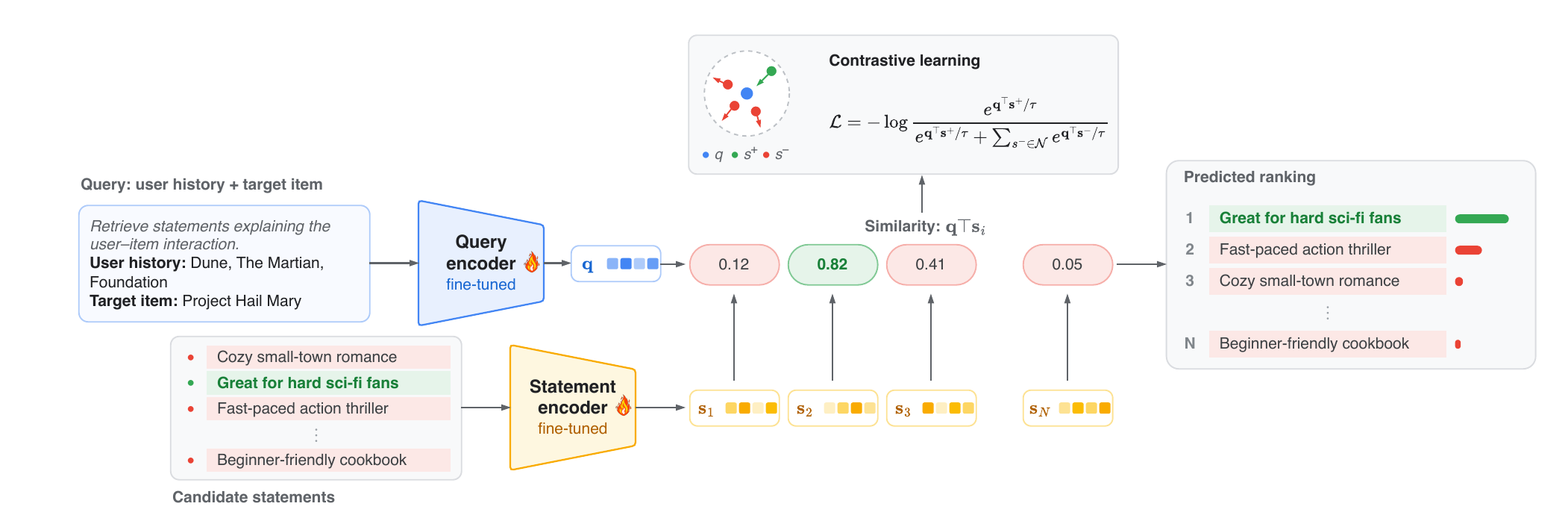}
\vspace{-6mm}
\caption{
Overview of \textsc{LMStaR}.
The query encoder represents the user's interaction history and target
item metadata, while the statement encoder independently encodes
candidate explanations.
Statements are ranked by cosine similarity in a shared embedding space.
Both encoders are jointly optimized through contrastive
learning.
}
\Description{
Overview of \textsc{LMStaR}.
The query encoder represents the user's interaction history and target
item metadata, while the statement encoder independently encodes
candidate explanations.
Statements are ranked by cosine similarity in a shared embedding space.
Both encoders are jointly optimized through contrastive
learning.
}
\label{fig:lmstar}
\end{figure*}

\section{Method}
\subsection{Preliminaries}
Explainable recommendation aims to justify why an item is recommended to a user based on their interaction history. This task can be formulated as a ranking problem, where candidate explanatory statements are scored and ranked according to their relevance to a given user--item interaction.

Let $\mathcal{U}$ and $\mathcal{I}$ denote the sets of users and items, respectively. Each item $i \in \mathcal{I}$ is associated with metadata $\mathcal{M}_i$, such as its name, category, and other descriptive attributes. Each user $u \in \mathcal{U}$ has a chronologically ordered interaction history $\mathcal{H}_u = (h_{u,1}, \ldots, h_{u,T_u})$, where $T_u$ denotes the number of observed interactions. An interaction with item $i$ is represented as $h_{u,t} = (\mathcal{M}_i, \mathcal{S}_{ui})$, where $\mathcal{S}_{ui} = \{s_1,\ldots,s_{n_{ui}}\}$ is the set of ground-truth explanatory statements describing the interaction between user $u$ and item $i$. We further define $\mathcal{S}_i = \bigcup_{u \in \mathcal{U}} \mathcal{S}_{ui}$ as the set of explanatory statements associated with item $i$.

The candidate statement set determines the ranking level. In \emph{global-level ranking}, candidates are drawn from the entire statement collection, whereas \emph{item-level ranking} restricts candidates to statements associated with the target item. We focus on item-level ranking, where the candidate set for an interaction $(u,i)$ is $\mathcal{S}_i$, with $\mathcal{S}_{ui} \subseteq \mathcal{S}_i$.

Given a user--item pair $(u,i)$, the objective is to learn a scoring function $f_\Theta$ that assigns a relevance score to each candidate statement $s \in \mathcal{S}_i$:
\begin{align}
    \hat{r}_{uis} &= f_\Theta(u,i,s), \notag \\
    \hat{\mathcal{S}}_{ui}^{\,k}
    &= \operatorname{TopK}_{s \in \mathcal{S}_i}
    \left(\hat{r}_{uis}\right),
\end{align}
where $\hat{\mathcal{S}}_{ui}^{\,k}$ denotes the $k$ highest-ranked statements presented as explanations for the recommendation. Ranking quality is evaluated against the ground-truth set $\mathcal{S}_{ui}$ using standard metrics such as nDCG@$k$~\cite{jarvelin2002cumulated}, Precision@$k$, and Recall@$k$.

\subsection{LMStaR: Language Models for Statement Ranking}

We introduce \textsc{LMStaR} (\textbf{L}anguage \textbf{M}odels for \textbf{Sta}tement \textbf{R}anking), a language-model-based approach to personalized explanation ranking that relies exclusively on textual representations of users, items, and candidate statements. We formulate the task as dense retrieval, where the user--item context serves as a query and candidate explanatory statements serve as documents. Figure~\ref{fig:lmstar} illustrates the overall architecture of \textsc{LMStaR}.

Our approach adopts a dual-encoder architecture comprising a query encoder $E^Q_\theta: \mathcal{X} \rightarrow \mathbb{R}^d$ and a statement encoder $E^S_\phi: \mathcal{X} \rightarrow \mathbb{R}^d$, where $\mathcal{X}$ denotes the space of textual inputs and $d$ is the embedding dimension. The query encoder represents a target user--item interaction through the user's interaction history $\mathcal{H}_u$ and the target item's metadata $\mathcal{M}_i$, while the statement encoder independently represents each candidate explanation. Both encoders map their respective inputs into a shared latent space, where relevance is measured through embedding similarity. Their parameters, $\Theta = \{\theta,\phi\}$, are jointly optimized through contrastive learning.

\subsubsection{User--Item Context Encoding}

The query encoder $E^Q_\theta$ maps the user--item context into a dense representation. To guide the encoding process toward personalized explanation retrieval, we employ the following instruction $P$:
\begin{quote}
\small\ttfamily
Retrieve explanation statements matching the target item to the user's preferences based on previous interactions.
\end{quote}
We construct the encoder input by combining this instruction with the user's interaction history $\mathcal{H}_u$ and the target item's metadata $\mathcal{M}_i$. The resulting query representation is defined as
\begin{align}
    \mathbf{q}_{ui}
    = E^Q_\theta(P, \mathcal{H}_u, \mathcal{M}_i),
\end{align}
where $\mathbf{q}_{ui} \in \mathbb{R}^d$ denotes the user--item context embedding. Specifically, we extract the final-layer hidden state of the last non-padding token and apply $\ell_2$ normalization. Under causal attention, this representation can aggregate information from all preceding tokens, encompassing the instruction, interaction history, and target item metadata. The query encoder is thus trained to capture the contextual information most relevant to identifying appropriate explanatory statements.

\subsubsection{Candidate Statement Encoding}

The statement encoder $E^S_\phi$ independently maps each candidate statement $s \in \mathcal{S}_i$ into the shared latent space:
\begin{align}
    \mathbf{s} = E^S_\phi(s),
\end{align}
where $\mathbf{s} \in \mathbb{R}^d$ denotes the statement embedding. Following the same pooling strategy as the query encoder, we extract the final-layer hidden state of the last non-padding token and apply $\ell_2$ normalization. 

\subsubsection{Relevance Scoring}

Given a target user--item pair $(u,i)$, we first compute its contextual representation $\mathbf{q}_{ui}$ using the query encoder $E^Q_\theta$. Each candidate statement $s \in \mathcal{S}_i$ is independently encoded into $\mathbf{s}$ using the statement encoder $E^S_\phi$. The relevance score is then computed as the dot product between their normalized embeddings:
\begin{align}
    \hat{r}_{uis}
    = \mathbf{q}_{ui}^{\top}\mathbf{s}.
\end{align}
Since both embeddings are $\ell_2$-normalized, the dot product is equivalent to cosine similarity, yielding $\hat{r}_{uis} \in [-1,1]$. Candidate statements are ranked in descending order of their relevance scores to identify the explanations that best match the user's preferences and the target item's characteristics.

\subsection{Model Training}

We jointly optimize the query and statement encoders to learn a shared embedding space in which cosine similarity serves as an effective relevance function for personalized explanation ranking. Specifically, we aim to bring user--item context embeddings closer to statements that explain the corresponding interactions, while pushing them away from irrelevant candidate statements.

\subsubsection{Training Data Construction}
We construct the training set from users' interaction histories while preserving their temporal order. For each interaction $h_{u,t}$ at time step $t$, we consider only the preceding interactions as the user's context, denoted by $\mathcal{H}_u^{<t} = (h_{u,1}, \ldots, h_{u,t-1})$. Consequently, the first interaction of each user is used exclusively to provide context for subsequent interactions, and training examples are constructed for $t = 2, \ldots, T_u$. For an interaction with item $i$, the query context is thus defined by $(\mathcal{H}_u^{<t}, \mathcal{M}_i)$.
We treat all ground-truth explanatory statements associated with the target interaction as positives, denoted by $\mathcal{P}_{ui} = \mathcal{S}_{ui}$. Under item-level ranking, the remaining statements associated with the target item constitute the negative candidate set $\mathcal{N}_{ui} = \mathcal{S}_i \setminus \mathcal{P}_{ui}$. In practice, we sample negatives from this set to limit computational cost, while retaining all positive statements. The resulting training set is defined as
\begin{align}
    \mathcal{D} = \big\{
    (\mathcal{H}_u^{<t}, \mathcal{M}_i,
    \mathcal{P}_{ui}, \mathcal{N}_{ui})
    \;\big|\;
    u \in \mathcal{U},\;
    t = 2,\ldots,T_u
    \big\},
\end{align}
where $i$ denotes the item involved in interaction $h_{u,t}$.

\subsubsection{Multi-Positive Contrastive Learning}
We adopt a contrastive objective based on InfoNCE~\cite{oord2018representation}, where each positive statement is independently contrasted against the negative statements of the corresponding interaction. The interaction-level loss is computed by averaging the individual positive-wise contrastive losses:
\begin{align}
    \mathcal{L}_{ui}
    = -\frac{1}{|\mathcal{P}_{ui}|}
    \sum_{s^+ \in \mathcal{P}_{ui}}
    \log
    \frac{
        \exp(\hat{r}_{uis^+}/\tau)
    }{
        \exp(\hat{r}_{uis^+}/\tau)
        + \sum_{s^- \in \mathcal{N}_{ui}}
        \exp(\hat{r}_{uis^-}/\tau)
    },
\end{align}
where $\tau$ is a temperature parameter. This formulation encourages each positive statement to receive a higher relevance score than the negatives, preventing well-ranked positives from compensating for poorly ranked ones.
The overall training objective averages the interaction-level losses:
\begin{align}
    \mathcal{L}(\mathcal{D};\theta,\phi)
    = \frac{1}{|\mathcal{D}|}
    \sum_{(u,i)\in\mathcal{D}}
    \mathcal{L}_{ui}.
\end{align}
The optimal parameters of the query and statement encoders are then obtained by minimizing this objective:
\begin{align}
    (\theta^*,\phi^*)
    = \arg\min_{\theta,\phi}
    \mathcal{L}(\mathcal{D};\theta,\phi).
\end{align}

\begin{table*}[ht]
\caption{Statement ranking results (\%). Best results in bold, second best \underline{underlined}. \textit{Ours}: \textsc{LMStaR} with the Qwen3-Embedding-8B backbone. \textit{Improv.}: absolute difference (points) with the best other method. $^\dagger$: significant difference between \textsc{LMStaR} and \textsc{CLER} (paired two-sided $t$-test over test users, $p<0.05$), marked on the better of the two.}
\vspace{-4mm}
\setlength{\tabcolsep}{4pt}
\begin{center}
\resizebox{\linewidth}{!}{
\begin{tabular}{lcccccccccccc}
\toprule
 & \multicolumn{4}{c}{\textbf{Beauty}} & \multicolumn{4}{c}{\textbf{Toys}} & \multicolumn{4}{c}{\textbf{Yelp}} \\
\cmidrule(lr){2-5} \cmidrule(lr){6-9} \cmidrule(lr){10-13}
\textbf{Methods} & \textbf{N@1} & \textbf{N@5} & \textbf{R@5} & \textbf{P@5} & \textbf{N@1} & \textbf{N@5} & \textbf{R@5} & \textbf{P@5} & \textbf{N@1} & \textbf{N@5} & \textbf{R@5} & \textbf{P@5} \\
\midrule
\multicolumn{13}{l}{\textit{Popularity-based}} \\
\textsc{ItemPop} & 2.628 & 4.123 & 5.813 & 2.069 & 2.033 & 3.769 & 5.779 & 1.910 & 1.190 & 1.681 & 2.229 & 0.745 \\
\textsc{GlobalPop} & 8.143 & 15.095 & 21.371 & 7.738 & 9.470 & 18.225 & 26.414 & 8.662 & 5.068 & 9.104 & 13.110 & 4.354 \\
\rowcolor{baselinebg}
\textsc{UserPop} & 10.179 & 16.975 & 23.476 & 8.630 & \underline{12.537} & 20.760 & 28.892 & 9.606 & 5.414 & 9.877 & 14.252 & 4.918 \\
\midrule
\multicolumn{13}{l}{\textit{Neighborhood-based}} \\
\textsc{RUCF} & 6.218 & 12.873 & 19.089 & 6.788 & 6.815 & 15.685 & 24.063 & 7.860 & 2.304 & 5.882 & 9.280 & 3.095 \\
\textsc{RICF} & 6.995 & 15.135 & 22.249 & 8.134 & 7.625 & 18.156 & 27.525 & 9.151 & 3.360 & 8.557 & 13.243 & 4.481 \\
\midrule
\multicolumn{13}{l}{\textit{Collaborative factorization}} \\
\textsc{CD} & 7.767 & 14.196 & 20.332 & 7.723 & 8.542 & 16.763 & 24.458 & 8.637 & 4.748 & 9.245 & 13.410 & 4.834 \\
\textsc{PITF} & 7.819 & 14.549 & 20.754 & 7.773 & 8.843 & 16.867 & 24.637 & 8.686 & 4.454 & 8.967 & 13.149 & 4.697 \\
\rowcolor{baselinebg}
\textsc{BPER} & 8.855 & 16.129 & 22.959 & 8.216 & 9.368 & 18.740 & 27.625 & 8.939 & 5.020 & 9.759 & 14.408 & 4.705 \\
\textsc{BPER-Large} & 7.871 & 13.693 & 19.596 & 7.334 & 10.093 & 17.016 & 24.156 & 8.220 & 3.148 & 6.427 & 9.748 & 3.311 \\
\midrule
\multicolumn{13}{l}{\textit{Semantic-enhanced}} \\
\textsc{BPER+} & 8.772 & 15.858 & 22.591 & 8.145 & 9.035 & 17.713 & 25.990 & 8.519 & 4.390 & 8.759 & 12.936 & 4.229 \\
\textsc{SE-BPER} & 8.476 & 14.329 & 20.188 & 7.604 & 10.618 & 17.689 & 24.948 & 8.485 & 3.494 & 6.829 & 10.299 & 3.513 \\
\textsc{SE-BPER+} & 8.491 & 14.296 & 20.246 & 7.619 & 10.563 & 17.304 & 24.367 & 8.334 & 3.366 & 6.784 & 10.174 & 3.541 \\
\midrule
\multicolumn{13}{l}{\textit{Contrastive learning}} \\
\rowcolor{baselinebg}
\textsc{CLER} & \underline{11.218} & \underline{19.741} & \underline{27.770} & \underline{10.384} & 12.521 & \underline{22.774} & \underline{32.551} & \underline{11.458} & \underline{8.965} & \textbf{16.508}$^\dagger$ & \textbf{23.544}$^\dagger$ & \textbf{8.498}$^\dagger$ \\
\midrule
\multicolumn{13}{l}{\textit{LLM-based neural retrieval}} \\
\rowcolor{baselinebg}
\textsc{Qwen3-Embedding-8B} \scriptsize (zero-shot) & 10.105 & 17.169 & 23.923 & 8.861 & 10.669 & 19.878 & 28.485 & 9.815 & 5.778 & 10.515 & 14.988 & 5.199 \\
\rowcolor{oursbg}
\textbf{\textsc{LMStaR} (Ours)} 
& \textbf{14.722}$^\dagger$ & \textbf{22.396}$^\dagger$ & \textbf{29.944}$^\dagger$ & \textbf{11.017}$^\dagger$ & \textbf{20.773}$^\dagger$ & \textbf{30.563}$^\dagger$ & \textbf{40.206}$^\dagger$ & \textbf{13.831}$^\dagger$ & \textbf{10.021}$^\dagger$ & \underline{15.406} & \underline{20.793} & \underline{7.345} \\
\rowcolor{oursbg}
\quad \textit{Improv.} & \textcolor{improvpos}{+3.504} & \textcolor{improvpos}{+2.655} & \textcolor{improvpos}{+2.174} & \textcolor{improvpos}{+0.633} & \textcolor{improvpos}{+8.236} & \textcolor{improvpos}{+7.789} & \textcolor{improvpos}{+7.655} & \textcolor{improvpos}{+2.373} & \textcolor{improvpos}{+1.056} & \textcolor{improvneg}{$-$1.102} & \textcolor{improvneg}{$-$2.751} & \textcolor{improvneg}{$-$1.153} \\
\bottomrule
\end{tabular}
}
\label{tab:statement-ranking-results-k5}
\end{center}
\vspace{-2mm}
\end{table*}

\section{Experiments}
We evaluate \textsc{LMStaR} on item-level statement ranking across three domains, comparing it with established baselines, examining backbone adaptation and cross-domain transfer, and analyzing the role of the query context.

\subsection{Experimental Setup}
\subsubsection{Datasets}
We conduct experiments on three real-world recommendation datasets: \textbf{Amazon Beauty} (\textit{Beauty}), \textbf{Amazon Toys} (\textit{Toys})~\cite{ni2019justifying}~\footnote{\url{https://jmcauley.ucsd.edu/data/amazon/links.html}}, and \textbf{Yelp}~\footnote{\url{https://www.yelp.com/dataset/challenge}}. As these datasets contain free-form user reviews rather than structured explanatory statements, we construct statement-level datasets through an extraction and clustering pipeline, followed by user-wise temporal splitting. 
Dataset construction, chronological splitting, and processed statistics are provided in Appendix~\ref{app:datasets}.

\subsubsection{Baselines}
We compare \textsc{LMStaR} against five groups of baselines: (1) \textbf{popularity-based methods}~\cite{kabongo2026rank}, including \textsc{ItemPop}, \textsc{GlobalPop}, and \textsc{UserPop}; (2) \textbf{neighborhood-based collaborative filtering methods}~\cite{li2021extra}, namely \textsc{RUCF} and \textsc{RICF}; (3) \textbf{factorization methods}, comprising (3a) \textit{purely collaborative models} such as \textsc{CD}~\cite{li2021extra}, \textsc{PITF}~\cite{li2021extra}, \textsc{BPER}~\cite{li2023relationship}, and \textsc{BPER-Large}~\cite{xu2022learning}, and (3b) \textit{semantically enhanced variants}, including \textsc{BPER+}~\cite{li2023relationship}, \textsc{SE-BPER}, and \textsc{SE-BPER+}~\cite{xu2022learning}; (4) c\textbf{ontrastive collaborative learning} with \textsc{CLER}~\cite{escarda2026contrastive}; and (5) \textbf{zero-shot neural retrieval} using \textsc{Qwen3-Embedding-8B}~\cite{zhang2025qwen3}. Detailed descriptions are provided in Appendix~\ref{app:baselines}.

\subsubsection{Evaluation Metrics}
We evaluate item-level statement ranking using Precision@$k$ (P@$k$), Recall@$k$ (R@$k$), and Normalized Discounted Cumulative Gain (N@$k$/NDCG@$k$)~\cite{jarvelin2002cumulated}, reporting results at $k \in \{1, 5\}$. Detailed definitions are provided in Appendix~\ref{app:metrics}.

\subsubsection{Model Details}
We evaluate several pretrained language-model backbones for \textsc{LMStaR}. Unless otherwise specified, we independently initialize the query and statement encoders from the official {Qwen3-Embedding-8B} checkpoint~\cite{zhang2025qwen3}
We empirically observe that complete fine-tuning can yield lower performance than parameter-efficient adaptation with LoRA~\cite{hu2022lora}. We therefore adopt LoRA to fine-tune both encoders by default. Further details on baseline implementations, model backbones, and training configurations are provided in Appendix~\ref{app:implementation}.

\subsection{Main Results}
We assess overall ranking effectiveness, robustness across pretrained backbones, parameter-efficient versus full fine-tuning, and cross-domain transferability.

\subsubsection{Overall Ranking Performance}
\label{sec:results-overall}

Table~\ref{tab:statement-ranking-results-k5} compares \textsc{LMStaR}
with existing statement ranking baselines on Beauty, Toys, and Yelp,
reporting NDCG@1, NDCG@5, Recall@5, and Precision@5.

\smallskip
\noindent\textbf{\textsc{LMStaR} achieves the highest first-rank
relevance across all domains, while its top-five effectiveness is
domain-dependent.}
\textsc{LMStaR} achieves the best results on 9 of the 12
dataset--metric combinations and ranks second on the remaining three.
Notably, it achieves the highest NDCG@1 on all three datasets,
significantly outperforming \textsc{CLER} by 3.50, 8.25, and 1.06
percentage points on Beauty, Toys, and Yelp, respectively
(31.2\%, 65.9\%, and 11.8\% relative improvements).
At $k=5$, \textsc{LMStaR} also significantly outperforms \textsc{CLER}
across all metrics on Beauty and Toys, improving NDCG@5 by 2.66
(+13.4\%) and 7.79 points (+34.2\%), respectively.
On Yelp, however, despite its significant first-rank advantage,
\textsc{LMStaR} trails \textsc{CLER} by 1.10 points in NDCG@5,
2.75 points in Recall@5, and 1.15 points in Precision@5.

\smallskip
\noindent\textbf{A pretrained text embedder provides competitive
zero-shot ranking, while task-specific adaptation yields substantial
additional gains.}
Without task-specific training, Qwen3-Embedding-8B outperforms all
neighborhood-based and factorization methods, including
semantic-enhanced variants, across all 12 dataset--metric combinations.
Adapting the same backbone through \textsc{LMStaR} improves all
evaluated metrics.
Specifically, NDCG@1 increases by 4.62 points (+45.7\%) on Beauty,
10.10 points (+94.7\%) on Toys, and 4.24 points (+73.4\%) on Yelp.
The corresponding NDCG@5 gains are 5.23 (+30.4\%), 10.69
(+53.8\%), and 4.89 points (+46.5\%), respectively.

\smallskip
\noindent\textbf{Semantic enhancements yield limited gains,
while popularity and \textsc{CLER} remain strong.}
None of the neighborhood-based or factorization methods, including
semantic-enhanced variants, surpasses \textsc{UserPop} in NDCG@1 or
NDCG@5 on any dataset, highlighting the competitiveness of
user-history frequency signals in item-level ranking~\cite{kabongo2026rank}.
Semantic enhancement yields mixed benefits: \textsc{BPER} outperforms
all its semantic-enhanced variants on NDCG@5, Recall@5, and
Precision@5 across datasets, although \textsc{SE-BPER} and
\textsc{SE-BPER+} achieve higher NDCG@1 on Toys and outperform
\textsc{BPER-Large} across all 12 dataset--metric combinations.
In contrast, \textsc{CLER}, trained with a contrastive NT-BXent
objective rather than BPR, outperforms all neighborhood-based,
factorization, and zero-shot retrieval baselines across all metrics,
as well as popularity baselines except on Toys N@1.

\begin{table}[t]
\caption{Statement ranking results (\%) of LMStaR backbones, N@5. \textsc{Zero-shot}: frozen backbone; \textsc{LoRA} / \textsc{Full FT}: LoRA / full fine-tuning of both encoders. $\Delta$: absolute difference (points) of full fine-tuning with the zero-shot backbone and with LoRA. Best results in bold, second best \underline{underlined}.}
\vspace{-4mm}
\begin{center}
\small
\setlength{\tabcolsep}{4pt}
\resizebox{\columnwidth}{!}{
\begin{tabular}{llccc}
\toprule
& & \textbf{Beauty} & \textbf{Toys} & \textbf{Yelp} \\
\midrule
\textbf{EmbeddingGemma-300M} 
&\textsc{Zero-shot} & 14.621 & 17.514 & 9.429 \\
&\textsc{LoRA} & 20.590 & \underline{28.985} & \underline{13.862} \\
&\textsc{Full FT} & 18.938 & 26.322 & 13.023 \\
&\quad $\Delta$ vs Zero-shot & \textcolor{improvpos}{+4.32} & \textcolor{improvpos}{+8.81} & \textcolor{improvpos}{+3.59} \\
&\quad $\Delta$ vs LoRA & \textcolor{improvneg}{$-$1.65}  &\textcolor{improvneg}{$-$2.66} & \textcolor{improvneg}{$-$0.84} \\
\midrule
\textbf{Qwen3-Embedding-0.6B}
&\textsc{Zero-shot} & 16.189 & 17.713 & 9.635 \\
&\textsc{LoRA} & \textbf{21.607} & \textbf{29.953} & \textbf{14.577} \\
&\textsc{Full FT} & \underline{20.609} & 28.936 & 13.454 \\
&\quad $\Delta$ vs Zero-shot & \textcolor{improvpos}{+4.42} & \textcolor{improvpos}{+11.22} & \textcolor{improvpos}{+3.82} \\
&\quad $\Delta$ vs LoRA & \textcolor{improvneg}{$-$1.00} & \textcolor{improvneg}{$-$1.02} & \textcolor{improvneg}{$-$1.12} \\
\bottomrule
\end{tabular}
}
\label{tab:lmstar-adaptation}
\end{center}
\vspace{-3mm}
\end{table}
\begin{figure}
\centering
\includegraphics[width=0.8\linewidth]{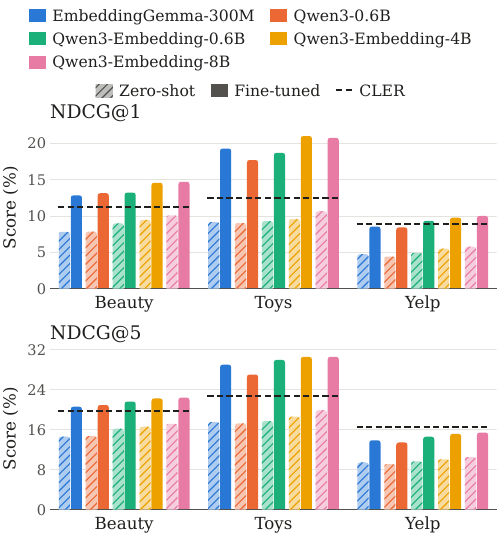}
\vspace{-4mm}
\caption{
Statement ranking results of LMStaR backbones (N@1 and N@5). \textsc{Zero-shot} (hatched): frozen backbone;
\textsc{Fine-tuned} (solid): task-specific adaptation. Dashed line: \textsc{CLER}, the strongest baseline.
}
\Description{
Two grouped bar charts stacked vertically (NDCG@1 above NDCG@5), one bar group per dataset (Beauty, Toys, Yelp). For each of five backbones, a hatched bar shows the frozen backbone and a solid bar the backbone after task-specific adaptation; a dashed line marks CLER on each dataset.
}
\label{fig:lmstar-backbones-ndcg}
\vspace{-4mm}
\end{figure}

\subsubsection{Backbone Robustness and Task-Specific Adaptation}
\label{sec:results-backbones}
In Figure~\ref{fig:lmstar-backbones-ndcg}, we compare the zero-shot
and LoRA-adapted performance of five pretrained backbones against
\textsc{CLER} in NDCG@1 and NDCG@5.
These include EmbeddingGemma~\cite{vera2025embeddinggemma} (300M),
Qwen3~\cite{yang2025qwen3} (0.6B), and
Qwen3-Embedding~\cite{zhang2025qwen3} (0.6B, 4B, and 8B).

\smallskip
\noindent\textbf{\textsc{LMStaR} generalizes across backbones,
with stronger relative first-rank performance.}
All five adapted backbones outperform \textsc{CLER} in both NDCG@1
and NDCG@5 on Beauty and Toys. Even EmbeddingGemma-300M surpasses
\textsc{CLER} by 0.85 points (+4.3\%) on Beauty and 6.21 points
(+27.3\%) on Toys in NDCG@5, showing that these advantages are
not specific to the 8B backbone.
On Yelp, the three adapted Qwen3-Embedding models outperform
\textsc{CLER} in NDCG@1 (9.35--10.02 vs.\ 8.97), whereas none
of the five backbones surpasses it in NDCG@5
(13.43--15.41 vs.\ 16.51).
At the same 0.6B scale, Qwen3-Embedding consistently outperforms
Qwen3 on both metrics, before and after adaptation, suggesting
a benefit from its retrieval-oriented, instruction-aware
embedding pretraining.

\smallskip
\noindent\textbf{Task-specific adaptation consistently improves
ranking effectiveness, exceeding the gains observed from
zero-shot model scaling.}
Adaptation improves NDCG@1 and NDCG@5 for every backbone on
all three datasets, yielding improvements in all 30 comparisons.
Moreover, every adapted backbone outperforms every zero-shot
backbone on both metrics across all three datasets, while
none of the zero-shot models surpasses \textsc{CLER}.
For instance, on Toys, adapted Qwen3-Embedding-0.6B exceeds
the zero-shot 8B model by 8.04 points (+75.3\%) in NDCG@1
and 10.08 points (+50.7\%) in NDCG@5.

\subsubsection{LoRA versus Full Fine-Tuning}
\label{sec:results-adaptation}

Table~\ref{tab:lmstar-adaptation} compares LoRA and full
fine-tuning of both encoders for EmbeddingGemma-300M and
Qwen3-Embedding-0.6B, reporting NDCG@5 on all three datasets.

\smallskip
\noindent\textbf{LoRA outperforms full fine-tuning across both
backbones and all three domains.}
Although full fine-tuning improves over zero-shot ranking
in all six configurations, it consistently underperforms
LoRA by 0.84--2.66 points.
The largest difference occurs on Toys with EmbeddingGemma-300M,
where LoRA achieves 28.99 NDCG@5 compared with 26.32 for full
fine-tuning, a gain of 2.66 points (+10.1\%).
These results support the use of parameter-efficient LoRA
adaptation in \textsc{LMStaR}, as updating all backbone
parameters does not yield better ranking effectiveness
in the evaluated settings.

\begin{table}[t]
\centering
\caption{
Cross-domain transfer of statement ranking in terms of N@1 and N@5.
Rows indicate the source (training) domain and columns the target (evaluation) domain.
The zero-shot backbone (Qwen3-Embedding-8B) and \textsc{CLER} (trained on the target domain) are given for reference.
}
\label{tab:transfer-ranking}
\small
\setlength{\tabcolsep}{3.5pt}

\vspace{-4mm}
\resizebox{\columnwidth}{!}{
\begin{tabular}{rccccccc}
\toprule

&
& \multicolumn{6}{c}{\textbf{Target}} \\

\cmidrule(lr){3-8}

&
& \multicolumn{2}{c}{\textbf{Beauty}}
& \multicolumn{2}{c}{\textbf{Toys}}
& \multicolumn{2}{c}{\textbf{Yelp}} \\

\cmidrule(lr){3-4}
\cmidrule(lr){5-6}
\cmidrule(lr){7-8}

\textbf{Model}
& \textbf{Source}
& \textbf{N@1} & \textbf{N@5}
& \textbf{N@1} & \textbf{N@5}
& \textbf{N@1} & \textbf{N@5} \\

\midrule

\rowcolor{baselinebg}
{Qwen3-Emb.-8B}
&
& 10.105 & 17.169 & 10.669 & 19.878 & 5.778 & 10.515 \\

\rowcolor{baselinebg}
\textsc{CLER}
&
& 11.218 & \underline{19.741} & 12.521 & 22.774 & \underline{8.965} & \textbf{16.508} \\

\midrule
\textbf{\textsc{LMStaR}}
& \textbf{Beauty} & \textbf{14.722} & \textbf{22.396} & \underline{18.529} & \underline{27.071} & 8.620 & 13.408 \\
& \textbf{Toys}   & \underline{12.520} & 19.624 & \textbf{20.773} & \textbf{30.563} & 7.967 & 12.586 \\
& \textbf{Yelp}   & 12.011 & 18.759 & 16.050 & 24.625 & \textbf{10.021} & \underline{15.406} \\

\bottomrule
\end{tabular}
}
\vspace{-3mm}
\end{table}

\subsubsection{Cross-Domain Transferability}
\label{sec:results-transfer}

Table~\ref{tab:transfer-ranking} evaluates the cross-domain
transferability of \textsc{LMStaR} using Qwen3-Embedding-8B.
We train on each source domain and evaluate directly on the
other two domains, without target-domain fine-tuning. We compare against the zero-shot backbone
and \textsc{CLER} trained on the target domain.

\smallskip
\noindent\textbf{Source-domain adaptation yields transferable ranking
gains, outperforming zero-shot and conventional target-domain baselines.}
All six transfer configurations outperform the zero-shot backbone
in both NDCG@1 and NDCG@5, yielding improvements across all
12 comparisons. They also surpass all popularity-based,
neighborhood-based, and factorization baselines evaluated on
the target domain (Table~\ref{tab:statement-ranking-results-k5}),
despite receiving no target-domain fine-tuning.
The largest gains over zero-shot ranking occur when transferring
from Beauty to Toys, improving NDCG@1 by 7.86 points (+73.7\%)
and NDCG@5 by 7.19 points (+36.2\%).

\smallskip
\noindent\textbf{Transferred \textsc{LMStaR} can outperform
target-trained \textsc{CLER}, but remains below in-domain adaptation.}
On Toys, both transferred models outperform \textsc{CLER}
in NDCG@1 and NDCG@5. Notably, Beauty-to-Toys transfer
achieves gains of 6.01 points (+48.0\%) in NDCG@1
and 4.30 points (+18.9\%) in NDCG@5 over \textsc{CLER},
despite being trained on a different domain.
On Beauty, both transfers exceed \textsc{CLER} at NDCG@1
but not at NDCG@5, whereas neither transfer surpasses
\textsc{CLER} on Yelp.
Nevertheless, in-domain \textsc{LMStaR} outperforms every
transferred configuration on both metrics. 

\subsection{Detailed Analysis}
We analyze the quantity, selection, and availability of user history, followed by counterfactual query perturbations. Unless stated otherwise, we use the validation-selected Qwen3-Embedding-8B checkpoint without further training; candidate statements, embeddings, and relevance labels remain fixed when modifying the query.

\label{sec:query-analysis}

\begin{figure}[t]
\centering
\includegraphics[width=0.75\linewidth]{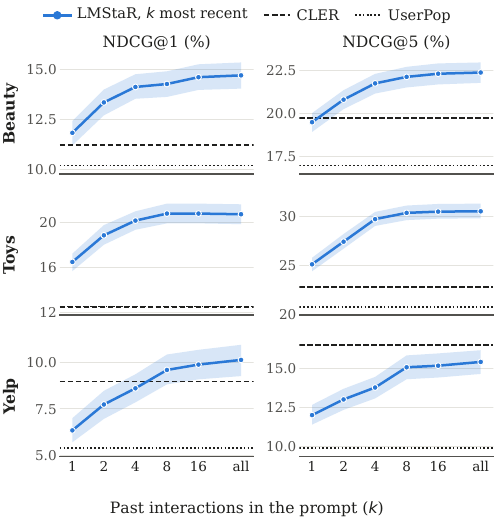}
\vspace{-4mm}
\caption{
LMStaR ranking performance with the $k$ most
recent past interactions.
Shaded bands: 95\% bootstrap confidence intervals.
}
\Description{
Six line charts, one row per dataset (Beauty, Toys, Yelp) and one column per metric (NDCG@1, NDCG@5). Each shows the score of LMStaR as the number of most recent past interactions in the prompt grows from 1 to 16 and to the whole history, with a shaded confidence band, and horizontal reference lines for CLER and UserPop.
}
\label{fig:history-amount}
\vspace{-4mm}
\end{figure}
\subsubsection{Impact of History Amount}
\label{sec:history-amount}

Figure~\ref{fig:history-amount} evaluates the effect of
retaining the $k \in \{1,2,4,8,16\}$ most recent past
interactions in the query.
Each retained interaction includes its item metadata and
associated statements; \emph{all} denotes the reference
prompt, containing as much recent history as fits within
the token budget.
We compare the resulting NDCG@1 and NDCG@5 scores
against \textsc{CLER} and \textsc{UserPop}, evaluated
on the same test users.

\smallskip
\noindent\textbf{Aggregate ranking gains are concentrated
in the first few recent interactions.}
From $k=1$ to \emph{all}, NDCG@5 rises from
19.5 to 22.4 on Beauty, 25.1 to 30.6 on Toys, and
12.0 to 15.4 on Yelp, with similar improvements in NDCG@1.
Notably, even with a single past interaction,
\textsc{LMStaR} exceeds \textsc{UserPop} on both metrics
across all datasets and \textsc{CLER} on Toys.
The aggregate gains become smaller at larger $k$,
particularly on Amazon.
However, this flattening partly reflects the distribution
of available history lengths (Table~\ref{tab:history-populations}).

\begin{figure}[t]
\centering
\includegraphics[width=0.85\linewidth]{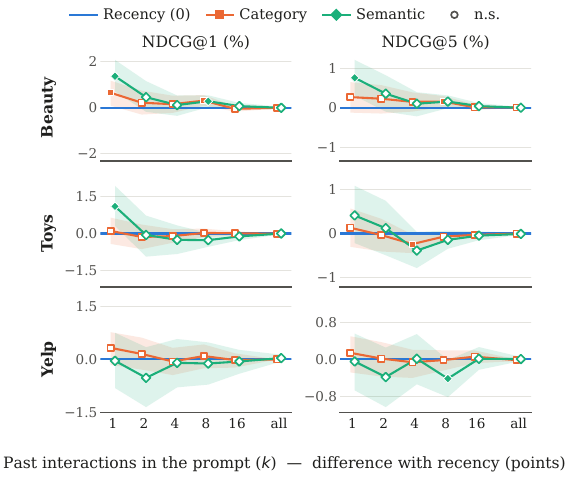}
\vspace{-4mm}
\caption{
Paired NDCG@1 and NDCG@5 differences (points)
between item-aware selection and recency. Shaded areas are 95\% paired bootstrap confidence intervals; filled markers are significant differences ($p<0.05$), hollow markers are not (n.s.).
}
\Description{
Six line charts, one row per dataset (Beauty, Toys, Yelp) and one column per metric (NDCG@1, NDCG@5). Each shows, for k from 1 to 16 and the whole history, the difference between selecting past interactions by category or semantic similarity with the target item and selecting the most recent ones, around a zero line standing for recency, with shaded confidence bands; filled markers indicate significant differences.
}
\label{fig:history-selection}
\vspace{-6mm}
\end{figure}
\subsubsection{History Selection Strategies}
\label{sec:history-selection}

Figure~\ref{fig:history-selection} compares \textit{recency-based}
history selection with two item-aware alternatives.
\emph{Category} prioritizes past items by the depth
of their shared category-path prefix with the target item,
whereas \emph{Semantic} ranks them by cosine similarity
between item-metadata embeddings from an independent,
non-adapted Qwen3-Embedding-0.6B model.
Ties favor more recent interactions.
We report paired NDCG differences relative to selecting
the $k$ most recent interactions.

\smallskip
\noindent\textbf{Recency remains a strong default, while
similarity-based selection offers its largest gains under
very small history budgets.}
Neither category-based nor semantic selection consistently
outperforms recency across datasets and history budgets.
The largest positive differences arise with a single
selected interaction: semantic selection improves
NDCG@1 by approximately 1.4 points on Beauty and
1.1 points on Toys, also improving NDCG@5 on Beauty
by roughly 0.7 points.
For larger budgets, differences are generally small and
mixed, partly because the selection policies often retain
the same interactions.
On Yelp, neither alternative provides a consistent gain.
Importantly, \textsc{LMStaR} was trained with recency-ordered
histories rather than these alternative selection policies.

\begin{figure}[t]
\centering
\includegraphics[width=0.8\linewidth]{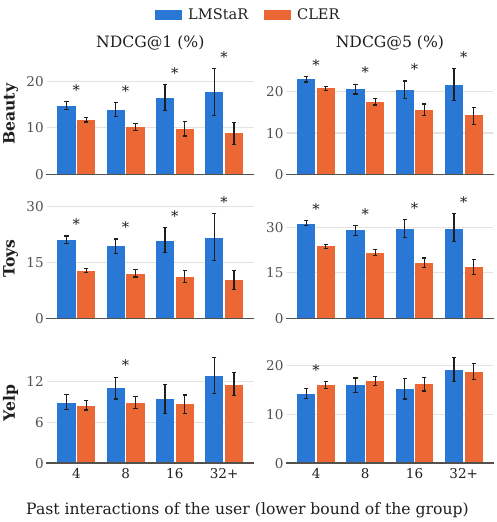}
\vspace{-4mm}
\caption{
LMStaR versus CLER across available
history-length groups (NDCG@1 and NDCG@5).
Error bars: 95\% bootstrap confidence intervals.
}
\Description{
Six grouped bar charts, one row per dataset (Beauty, Toys, Yelp) and one column per metric (NDCG@1, NDCG@5). For each group of test users by number of past interactions (4-7, 8-15, 16-31, 32 or more), a bar shows LMStaR and a bar shows CLER, with confidence intervals; stars mark significant differences.
}
\label{fig:history-groups}
\vspace{-4mm}
\end{figure}
\subsubsection{Performance across History-Length Groups}
\label{sec:history-groups}

Figure~\ref{fig:history-groups} examines how ranking
performance varies with the amount of history available
for each test user.
We partition users into four groups (4--7, 8--15,
16--31, and at least 32 past interactions), with
populations reported in Table~\ref{tab:history-populations}.
Unlike the preceding interventions, this analysis uses
the original test predictions rather than modifying
the query.
We compare \textsc{LMStaR} with \textsc{CLER} on the
same interactions, averaging \textsc{CLER}'s
per-interaction metrics across its three runs.

\smallskip
\noindent\textbf{\textsc{LMStaR} consistently outperforms
\textsc{CLER} across history-length groups on Amazon,
while its top-five disadvantage on Yelp is concentrated
among users with shorter histories.}
On Beauty and Toys, \textsc{LMStaR} achieves higher
NDCG@1 and NDCG@5 in all four groups, with numerically
larger advantages among users with longer available
histories.
This pattern mainly reflects lower groupwise scores
for \textsc{CLER}, whereas \textsc{LMStaR} maintains
comparatively stable ranking effectiveness.
Yelp exhibits a different pattern: \textsc{CLER}'s
clearest NDCG@5 advantage occurs in the 4--7 group,
which accounts for 49.2\% of test users, while
the differences become smaller or reverse numerically
in groups with longer histories.
These comparisons are observational, however:
the groups differ in more than history length, and
available interactions may exceed the number actually
retained within the query token budget.

\begin{table}[t]
\caption{Counterfactual queries (N@5, \%; $\Delta$: difference with the reference, paired bootstrap). Candidates and ground truth are unchanged; only the query is modified. $^\dagger$Out of distribution (never seen in training). 
}
\vspace{-4mm}
\begin{center}
\setlength{\tabcolsep}{4pt}
\resizebox{\linewidth}{!}{
\begin{tabular}{lcccccc}
\toprule
 & \multicolumn{2}{c}{\textbf{Beauty}} & \multicolumn{2}{c}{\textbf{Toys}} & \multicolumn{2}{c}{\textbf{Yelp}} \\
\cmidrule(lr){2-3} \cmidrule(lr){4-5} \cmidrule(lr){6-7}
\textbf{Query} & N@5 & $\Delta$ & N@5 & $\Delta$ & N@5 & $\Delta$ \\
\midrule
\rowcolor{oursbg}
Reference (own history + item) & 22.40 & -- & 30.56 & -- & 15.41 & -- \\
\midrule
\rowcolor{baselinebg}
(a) Empty history$^\dagger$ & 16.30 & $-$6.09 & 18.26 & $-$12.30 & 10.56 & $-$4.84 \\
(b) Empty item$^\dagger$ & 21.59 & $-$0.80 & 30.18 & $-$0.38 & 15.35 & $-$0.06 \\
(c) Item of the same category & 21.62 & $-$0.78 & 30.11 & $-$0.51 & 15.25 & $-$0.13 \\
(d) Random item & 21.43 & $-$0.97 & 30.16 & $-$0.41 & 15.05 & $-$0.36 \\
\rowcolor{baselinebg}
(e) History of another user & 15.66 & $-$6.73 & 17.36 & $-$13.20 & 9.76 & $-$5.65 \\
\bottomrule
\end{tabular}
}
\label{tab:probes}
\end{center}
\vspace{-4mm}
\end{table}
\subsubsection{Counterfactual Query Analysis}
\label{sec:results-counterfactual}

Specifically, we remove the user history (a) or target-item
metadata (b), replace the target item with one from the same
category (c) or a random item (d), or substitute another user's
history matched on the number of past interactions (e).
Probes (a) and (b) introduce empty fields unseen during training,
whereas (c)--(e) preserve valid input fields.

\smallskip
\noindent\textbf{\textsc{LMStaR} strongly relies on user-specific
interaction histories for personalized statement ranking.}
Replacing the user's history with that of another user (e) reduces NDCG@5 by 6.73 points (30.0\%) on Beauty,
13.20 points (43.2\%) on Toys, and 5.65 points (36.7\%) on Yelp.
Removing the history entirely (a) produces similarly large
degradations of 6.09, 12.30, and 4.84 points, respectively,
although this configuration is unseen during training.
Notably, with another user's history (e), \textsc{LMStaR} falls
below \textsc{UserPop} and its own zero-shot backbone on all
three datasets (Table~\ref{tab:statement-ranking-results-k5}).

\smallskip
\noindent\textbf{Target-item metadata has a comparatively limited
influence under item-level ranking.}
In contrast, replacing the target item with a random item (d)
reduces NDCG@5 by only 0.97 points (4.3\%) on Beauty,
0.41 points (1.3\%) on Toys, and 0.36 points (2.3\%) on Yelp.
The much smaller sensitivity to item modifications may
partly reflect the item-level evaluation setting, where
candidate statements are already restricted to the
actual target item.

\section{Conclusion}
\label{sec:conclusion}

We introduced \textsc{LMStaR}, a text-based dual-encoder
framework that ranks explanatory statements using a query
combining user history and target-item metadata, with both
encoders adapted through LoRA and multi-positive contrastive
learning. On Beauty, Toys, and Yelp, \textsc{LMStaR} achieves
the highest NDCG@1, and leads the top-five metrics on both
Amazon datasets, while \textsc{CLER} remains stronger at
$k=5$ on Yelp. Task-specific adaptation improves all five
evaluated backbones, with smaller adapted models outperforming
larger zero-shot alternatives; LoRA also surpasses full
fine-tuning for the two tested backbones. Adaptation transfers
across domains without target-domain fine-tuning.
History analyses reveal substantial gains from a few recent
interactions and, critically, a strong dependence on the
user's own history. Together, these findings establish
task-adapted language-model representations as a promising
approach to personalized explanation ranking.

\section{Discussion}
\label{sec:discussion}

The contrast between first-rank relevance and top-five
coverage on Yelp raises an important question: a single
query vector may struggle to represent multiple explanatory
aspects, while \textsc{CLER}'s collaborative signals may
provide complementary evidence. Neither mechanism is
established by our experiments, but both motivate hybrid
or multi-vector ranking approaches. Counterfactual probes
also show much greater sensitivity to user-history
substitutions than to changes in the target-item description.
This does not imply that item information is irrelevant:
item-level evaluation already restricts candidates to
the true target item, potentially reducing the marginal
contribution of its metadata.
The evaluation also has limitations. Binary relevance
is derived from automatically extracted and clustered
review statements; unobserved statements may still be
relevant, and ranking metrics do not establish perceived
usefulness or faithfulness. Dataset construction relies
on Qwen-family models, potentially introducing shared
biases, although gains with EmbeddingGemma show that
adaptation is not restricted to Qwen backbones.
Importantly, item-level ranking targets
personalization more directly than global
retrieval, but candidate pools may
include statements from held-out reviews unavailable at
recommendation time, and test items are restricted to
those observed during training. More realistic evaluations
should consider temporally available candidate pools,
sparsely reviewed or unseen items, and human assessments
of explanation quality.

\section*{Acknowledgements}
This work was granted access to the HPC resources of IDRIS under the allocation 2026-AD011016963R1 made by GENCI.

\bibliographystyle{ACM-Reference-Format}
\bibliography{output}

@inproceedings{kabongo2026rank,
author = {Kabongo, Ben and Satouf, Arthur and Guigue, Vincent},
title = {Rank, Don't Generate: Statement-level Ranking for Explainable Recommendation},
year = {2026},
isbn = {9798400726002},
publisher = {Association for Computing Machinery},
address = {New York, NY, USA},
url = {https://doi.org/10.1145/3805713.3820414},
doi = {10.1145/3805713.3820414},
booktitle = {Proceedings of the 2026 International ACM SIGIR Conference on Innovative Concepts and Theories in Information Retrieval (ICTIR)},
pages = {159–170},
numpages = {12},
location = {Melbourne, VIC, Australia},
series = {ICTIR '26}
}

@article{jarvelin2002cumulated,
  title={Cumulated gain-based evaluation of IR techniques},
  author={J{\"a}rvelin, Kalervo and Kek{\"a}l{\"a}inen, Jaana},
  journal={ACM Transactions on Information Systems (TOIS)},
  volume={20},
  number={4},
  pages={422--446},
  year={2002},
  publisher={ACM New York, NY, USA}
}

@inproceedings{li2021extra,
  title={Extra: Explanation ranking datasets for explainable recommendation},
  author={Li, Lei and Zhang, Yongfeng and Chen, Li},
  booktitle={Proceedings of the 44th International ACM SIGIR conference on Research and Development in Information Retrieval},
  pages={2463--2469},
  year={2021}
}

@article{li2023relationship,
  title={On the relationship between explanation and recommendation: Learning to rank explanations for improved performance},
  author={Li, Lei and Zhang, Yongfeng and Chen, Li},
  journal={ACM Transactions on Intelligent Systems and Technology},
  volume={14},
  number={2},
  pages={1--24},
  year={2023},
  publisher={ACM New York, NY}
}

@article{xu2022learning,
  title={Learning to rank rationales for explainable recommendation},
  author={Xu, Zhichao and Han, Yi and Yang, Tao and Tran, Anh and Ai, Qingyao},
  journal={arXiv preprint arXiv:2206.05368},
  year={2022}
}

@article{escarda2026contrastive,
  title={Contrastive Learning for Explanation Ranking},
  author={Escarda-Fern{\'a}ndez, Miguel and Cancela, Brais and Eiras-Franco, Carlos and Guijarro-Berdi{\~n}as, Bertha and Alonso-Betanzos, Amparo},
  journal={Machine Learning},
  volume={115},
  number={7},
  pages={156},
  year={2026},
  publisher={Springer}
}

@inproceedings{li2020generate,
  title={Generate neural template explanations for recommendation},
  author={Li, Lei and Zhang, Yongfeng and Chen, Li},
  booktitle={Proceedings of the 29th ACM International Conference on Information \& Knowledge Management},
  pages={755--764},
  year={2020}
}

@inproceedings{dong2017learning,
  title={Learning to generate product reviews from attributes},
  author={Dong, Li and Huang, Shaohan and Wei, Furu and Lapata, Mirella and Zhou, Ming and Xu, Ke},
  booktitle={Proceedings of the 15th Conference of the European Chapter of the Association for Computational Linguistics: Volume 1, Long Papers},
  pages={623--632},
  year={2017}
}

@inproceedings{ni2019justifying,
  title={Justifying recommendations using distantly-labeled reviews and fine-grained aspects},
  author={Ni, Jianmo and Li, Jiacheng and McAuley, Julian},
  booktitle={Proceedings of the 2019 conference on empirical methods in natural language processing and the 9th international joint conference on natural language processing (EMNLP-IJCNLP)},
  pages={188--197},
  year={2019}
}

@inproceedings{li2017neural,
  title={Neural rating regression with abstractive tips generation for recommendation},
  author={Li, Piji and Wang, Zihao and Ren, Zhaochun and Bing, Lidong and Lam, Wai},
  booktitle={Proceedings of the 40th International ACM SIGIR conference on Research and Development in Information Retrieval},
  pages={345--354},
  year={2017}
}

@inproceedings{li2021personalized,
  title={Personalized transformer for explainable recommendation},
  author={Li, Lei and Zhang, Yongfeng and Chen, Li},
  booktitle={Proceedings of the 59th Annual Meeting of the Association for Computational Linguistics and the 11th International Joint Conference on Natural Language Processing (Volume 1: Long Papers)},
  pages={4947--4957},
  year={2021}
}

@article{li2023personalized,
  title={Personalized prompt learning for explainable recommendation},
  author={Li, Lei and Zhang, Yongfeng and Chen, Li},
  journal={ACM Transactions on Information Systems},
  volume={41},
  number={4},
  pages={1--26},
  year={2023},
  publisher={ACM New York, NY, USA}
}

@inproceedings{ma2024xrec,
  title={Xrec: Large language models for explainable recommendation},
  author={Ma, Qiyao and Ren, Xubin and Huang, Chao},
  booktitle={Findings of the Association for Computational Linguistics: EMNLP 2024},
  pages={391--402},
  year={2024}
}

@inproceedings{li2025g,
  title={G-refer: Graph retrieval-augmented large language model for explainable recommendation},
  author={Li, Yuhan and Zhang, Xinni and Luo, Linhao and Chang, Heng and Ren, Yuxiang and King, Irwin and Li, Jia},
  booktitle={Proceedings of the ACM on Web Conference 2025},
  pages={240--251},
  year={2025}
}

@inproceedings{reimers2019sentence,
  title={Sentence-bert: Sentence embeddings using siamese bert-networks},
  author={Reimers, Nils and Gurevych, Iryna},
  booktitle={Proceedings of the 2019 conference on empirical methods in natural language processing and the 9th international joint conference on natural language processing (EMNLP-IJCNLP)},
  pages={3982--3992},
  year={2019}
}

@inproceedings{karpukhin2020dense,
  title={Dense passage retrieval for open-domain question answering},
  author={Karpukhin, Vladimir and Oguz, Barlas and Min, Sewon and Lewis, Patrick and Wu, Ledell and Edunov, Sergey and Chen, Danqi and Yih, Wen-tau},
  booktitle={Proceedings of the 2020 conference on empirical methods in natural language processing (EMNLP)},
  pages={6769--6781},
  year={2020}
}

@inproceedings{xiong2020approximate,
  author={Xiong, Lee and Xiong, Chenyan and Li, Ye and Tang, Kwok-Fung and Liu, Jialin and Bennett, Paul and Ahmed, Junaid and Overwijk, Arnold},
  title = {Approximate Nearest Neighbor Negative Contrastive Learning for Dense Text Retrieval},
  booktitle = {International Conference on Learning Representations (ICLR)},
  year = {2021},
  url = {https://openreview.net/forum?id=zeFrfgyZln}
}

@article{wang2022text,
  title={Text embeddings by weakly-supervised contrastive pre-training},
  author={Wang, Liang and Yang, Nan and Huang, Xiaolong and Jiao, Binxing and Yang, Linjun and Jiang, Daxin and Majumder, Rangan and Wei, Furu},
  journal={arXiv preprint arXiv:2212.03533},
  year={2022}
}

@inproceedings{su2023one,
  title={One embedder, any task: Instruction-finetuned text embeddings},
  author={Su, Hongjin and Shi, Weijia and Kasai, Jungo and Wang, Yizhong and Hu, Yushi and Ostendorf, Mari and Yih, Wen-tau and Smith, Noah A and Zettlemoyer, Luke and Yu, Tao},
  booktitle={Findings of the Association for Computational Linguistics: ACL 2023},
  pages={1102--1121},
  year={2023}
}

@inproceedings{ma2024fine,
  title={Fine-tuning llama for multi-stage text retrieval},
  author={Ma, Xueguang and Wang, Liang and Yang, Nan and Wei, Furu and Lin, Jimmy},
  booktitle={Proceedings of the 47th International ACM SIGIR Conference on Research and Development in Information Retrieval},
  pages={2421--2425},
  year={2024}
}

@inproceedings{behnamghader2024llm2vec,
  author={BehnamGhader, Parishad and Adlakha, Vaibhav and Mosbach, Marius and Bahdanau, Dzmitry and Chapados, Nicolas and Reddy, Siva},
  title = {{LLM2Vec}: Large Language Models Are Secretly Powerful Text Encoders},
  booktitle = {Conference on Language Modeling (COLM)},
  year = {2024},
  url = {https://openreview.net/forum?id=IW1PR7vEBf}
}

@inproceedings{covington2016deep,
  title={Deep neural networks for youtube recommendations},
  author={Covington, Paul and Adams, Jay and Sargin, Emre},
  booktitle={Proceedings of the 10th ACM conference on recommender systems},
  pages={191--198},
  year={2016}
}

@inproceedings{hou2022towards,
  title={Towards universal sequence representation learning for recommender systems},
  author={Hou, Yupeng and Mu, Shanlei and Zhao, Wayne Xin and Li, Yaliang and Ding, Bolin and Wen, Ji-Rong},
  booktitle={Proceedings of the 28th ACM SIGKDD conference on knowledge discovery and data mining},
  pages={585--593},
  year={2022}
}

@inproceedings{li2023text,
  title={Text is all you need: Learning language representations for sequential recommendation},
  author={Li, Jiacheng and Wang, Ming and Li, Jin and Fu, Jinmiao and Shen, Xin and Shang, Jingbo and McAuley, Julian},
  booktitle={Proceedings of the 29th ACM SIGKDD conference on knowledge discovery and data mining},
  pages={1258--1267},
  year={2023}
}

@article{xu2026case,
  title={The Case Against Generation for Retrieval: Discriminative Language Models as Effective Retrievers},
  author={Xu, Zhe and Agrawal, Prachi and Asadi, Kavosh and Chen, Tianyi and Hu, Carl and Johnson, Justin and Lan, Wuwei and Liang, Mingfu and Liu, Xi and Lui, Tik On and others},
  journal={arXiv preprint arXiv:2607.25346},
  year={2026}
}

@inproceedings{hu2022lora,
  title     = {Lo{RA}: Low-Rank Adaptation of Large Language Models},
  author={Hu, Edward J and Shen, Yelong and Wallis, Phillip and Allen-Zhu, Zeyuan and Li, Yuanzhi and Wang, Shean and Wang, Lu and Chen, Weizhu},
  booktitle = {International Conference on Learning Representations (ICLR)},
  year      = {2022},
  url       = {https://openreview.net/forum?id=nZeVKeeFYf9}
}

@article{vera2025embeddinggemma,
  title={Embeddinggemma: Powerful and lightweight text representations},
  author={Vera, Henrique Schechter and Dua, Sahil and Zhang, Biao and Salz, Daniel and Mullins, Ryan and Panyam, Sindhu Raghuram and Smoot, Sara and Naim, Iftekhar and Zou, Joe and Chen, Feiyang and others},
  journal={arXiv preprint arXiv:2509.20354},
  year={2025}
}

@article{yang2025qwen3,
  title={Qwen3 technical report},
  author={Yang, An and Li, Anfeng and Yang, Baosong and Zhang, Beichen and Hui, Binyuan and Zheng, Bo and Yu, Bowen and Gao, Chang and Huang, Chengen and Lv, Chenxu and others},
  journal={arXiv preprint arXiv:2505.09388},
  year={2025}
}

@article{zhang2025qwen3,
  title={Qwen3 Embedding: Advancing Text Embedding and Reranking Through Foundation Models},
  author={Zhang, Yanzhao and Li, Mingxin and Long, Dingkun and Zhang, Xin and Lin, Huan and Yang, Baosong and Xie, Pengjun and Yang, An and Liu, Dayiheng and Lin, Junyang and others},
  journal={arXiv preprint arXiv:2506.05176},
  year={2025}
}

@misc{qwen38,
    title = {{Qwen3.8-Max}: A New Bar for Coding and Cowork},
    url = {https://qwen.ai/blog?id=qwen3.8},
    author = {{Qwen Team}},
    month = {August},
    year = {2026}
}

@inproceedings{
zhang2019bertscore,
title={BERTScore: Evaluating Text Generation with BERT},
author={Tianyi Zhang* and Varsha Kishore* and Felix Wu* and Kilian Q. Weinberger and Yoav Artzi},
booktitle={International Conference on Learning Representations},
year={2020},
url={https://openreview.net/forum?id=SkeHuCVFDr}
}

@inproceedings{sellam2020bleurt,
  title={BLEURT: Learning robust metrics for text generation},
  author={Sellam, Thibault and Das, Dipanjan and Parikh, Ankur},
  booktitle={Proceedings of the 58th annual meeting of the association for computational linguistics},
  pages={7881--7892},
  year={2020}
}

@inproceedings{papineni2002bleu,
  title={Bleu: a method for automatic evaluation of machine translation},
  author={Papineni, Kishore and Roukos, Salim and Ward, Todd and Zhu, Wei-Jing},
  booktitle={Proceedings of the 40th annual meeting of the Association for Computational Linguistics},
  pages={311--318},
  year={2002}
}

@inproceedings{lin2004rouge,
  title={Rouge: A package for automatic evaluation of summaries},
  author={Lin, Chin-Yew},
  booktitle={Text summarization branches out},
  pages={74--81},
  year={2004}
}

@article{rendle2012bpr,
  title={BPR: Bayesian personalized ranking from implicit feedback},
  author={Rendle, Steffen and Freudenthaler, Christoph and Gantner, Zeno and Schmidt-Thieme, Lars},
  journal={arXiv preprint arXiv:1205.2618},
  year={2012}
}

@article{oord2018representation,
  title={Representation learning with contrastive predictive coding},
  author={Oord, Aaron van den and Li, Yazhe and Vinyals, Oriol},
  journal={arXiv preprint arXiv:1807.03748},
  year={2018}
}

@inproceedings{kabongo2025factual,
  title={On the Factual Consistency of Text-based Explainable Recommendation Models},
  author={Kabongo, Ben and Guigue, Vincent},
  booktitle={Pacific-Asia Conference on Knowledge Discovery and Data Mining},
  pages={16--28},
  year={2026},
  organization={Springer}
}

@article{fu2023gptscore,
  title={Gptscore: Evaluate as you desire},
  author={Fu, Jinlan and Ng, See-Kiong and Jiang, Zhengbao and Liu, Pengfei},
  journal={arXiv preprint arXiv:2302.04166},
  year={2023}
}

@inproceedings{devlin2019bert,
  title={Bert: Pre-training of deep bidirectional transformers for language understanding},
  author={Devlin, Jacob and Chang, Ming-Wei and Lee, Kenton and Toutanova, Kristina},
  booktitle={Proceedings of the 2019 conference of the North American chapter of the association for computational linguistics: human language technologies, volume 1 (long and short papers)},
  pages={4171--4186},
  year={2019}
}

@inproceedings{sheng2025analyzing,
  title={Analyzing Uncertainty of LLM-as-a-Judge: Interval Evaluations with Conformal Prediction},
  author={Sheng, Huanxin and Liu, Xinyi and He, Hangfeng and Zhao, Jieyu and Kang, Jian},
  booktitle={Proceedings of the 2025 Conference on Empirical Methods in Natural Language Processing},
  pages={11297--11339},
  year={2025}
}

@article{zhang2020explainable,
  title={Explainable recommendation: A survey and new perspectives},
  author={Zhang, Yongfeng and Chen, Xu and others},
  journal={Foundations and Trends{\textregistered} in Information Retrieval},
  volume={14},
  number={1},
  pages={1--101},
  year={2020},
  publisher={Now Publishers, Inc.}
}

@book{anand2011mining,
  title={Mining of massive datasets},
  author={Anand, Rajaraman and Jeffrey David, Ullman},
  year={2011},
  publisher={Cambridge university press}
}

@inproceedings{huang2013learning,
  title={Learning deep structured semantic models for web search using clickthrough data},
  author={Huang, Po-Sen and He, Xiaodong and Gao, Jianfeng and Deng, Li and Acero, Alex and Heck, Larry},
  booktitle={Proceedings of the 22nd ACM international conference on Information \& Knowledge Management},
  pages={2333--2338},
  year={2013}
}

@inproceedings{stolfo2024groundedness,
    title = "Groundedness in Retrieval-augmented Long-form Generation: An Empirical Study",
    author = "Stolfo, Alessandro",
    editor = "Duh, Kevin  and
      Gomez, Helena  and
      Bethard, Steven",
    booktitle = "Findings of the {A}ssociation for {C}omputational {L}inguistics: {NAACL} 2024",
    month = jun,
    year = "2024",
    address = "Mexico City, Mexico",
    publisher = "Association for Computational Linguistics",
    url = "https://aclanthology.org/2024.findings-naacl.100/",
    doi = "10.18653/v1/2024.findings-naacl.100",
    pages = "1537--1552"
}

@inproceedings{qin2025meta,
    title = "Meta-Tool: Unleash Open-World Function Calling Capabilities of General-Purpose Large Language Models",
    author = "Qin, Shengqian  and
      Zhu, Yakun  and
      Mu, Linjie  and
      Zhang, Shaoting  and
      Zhang, Xiaofan",
    editor = "Che, Wanxiang  and
      Nabende, Joyce  and
      Shutova, Ekaterina  and
      Pilehvar, Mohammad Taher",
    booktitle = "Proceedings of the 63rd Annual Meeting of the {A}ssociation for {C}omputational {L}inguistics (Volume 1: Long Papers)",
    month = jul,
    year = "2025",
    address = "Vienna, Austria",
    publisher = "Association for Computational Linguistics",
    url = "https://aclanthology.org/2025.acl-long.1481/",
    doi = "10.18653/v1/2025.acl-long.1481",
    pages = "30653--30677",
    ISBN = "979-8-89176-251-0"
}

@String{Computing = "Computing" }

@String{Computer = "{IEEE} Computer" }

@String{Springer = "Springer-Verlag" }

@BOOK{test,
   author = "Donald E. Knuth",
   title = "Seminumerical Algorithms",
   volume = 2,
   series = "The Art of Computer Programming",
   publisher = "Addison-Wesley",
   address = "Reading, MA",
   edition = "2nd",
   month = "10~" # jan,
   year = "1981",
}

@ArtifactSoftware{R,
    title = {R: A Language and Environment for Statistical Computing},
    author = {{R Core Team}},
    organization = {R Foundation for Statistical Computing},
    address = {Vienna, Austria},
    year = {2019},
    url = {https://www.R-project.org/},
}

\appendix
\section{Datasets}
\label{app:datasets}
We conduct experiments on three real-world recommendation datasets: \textbf{Amazon Beauty} (\textit{Beauty}), \textbf{Amazon Toys} (\textit{Toys}), and \textbf{Yelp}. We first describe the construction of statement-level datasets from user reviews through statement extraction and clustering, followed by filtering and data splitting. Table~\ref{tab:dataset_stats} summarizes the resulting dataset statistics.

\paragraph{Dataset Construction.}
We build upon the statement extraction and clustering pipeline introduced by \textsc{StaR}~\cite{kabongo2026rank}, with two refinements aimed at improving the explanatory relevance and semantic consistency of the resulting statements.
First, we revise the extraction prompt to explicitly identify statements that explain the match between a user's preferences and an item's characteristics, rather than merely describing item attributes. Second, we strengthen the final clustering stage with reranker-based verification to reduce erroneous merges between semantically distinct clusters. Specifically, we select $q$ representative statements from each cluster to capture its semantic diversity in the embedding space. For each pair of candidate clusters, we evaluate all cross-cluster pairs of representatives and merge the clusters only if the reranker identifies every pair as semantically equivalent.
We use {Qwen3.8-27B}~\cite{qwen38}\footnote{\url{https://huggingface.co/Qwen/Qwen3.8-27B}} for statement extraction, {Qwen3-Embedding-0.6B}~\cite{zhang2025qwen3}\footnote{\url{https://huggingface.co/Qwen/Qwen3-Embedding-0.6B}} for semantic representations, and {Qwen3-Reranker-0.6B}~\cite{zhang2025qwen3}\footnote{\url{https://huggingface.co/Qwen/Qwen3-Reranker-0.6B}} for pairwise paraphrase verification. All prompts and hyperparameters are provided in the accompanying code.

\paragraph{Filtering and Data Splitting.}
After statement extraction and clustering, we discard interactions without valid explanatory statements and apply iterative $5$-core filtering, retaining only users and items with at least five interactions. We then perform a user-wise chronological split, holding out each user's most recent interaction for testing and the second most recent for validation. All earlier interactions form the training set. To ensure comparability with collaborative filtering baselines, we exclude test interactions involving items absent from the training set, thereby avoiding item cold-start evaluation. All methods are evaluated using the same data splits and item-level candidate statement sets.

\begin{table}[t]
\centering
\caption{Statistics of the statement-ranking datasets.}
\label{tab:dataset_stats}
\vspace{-4mm}
\resizebox{\linewidth}{!}{
\begin{tabular}{@{}lrrr@{}}

\toprule
& \textbf{Beauty} 
& \textbf{Toys} 
& \textbf{Yelp} \\

\midrule
\textbf{Dataset} \\
\;\; Users 
& 10\,819 
& 8\,556 
& 5\,209 \\

\;\; Items 
& 6\,289 
& 5\,662 
& 3\,787 \\

\;\; Interactions 
& 97\,484 
& 74\,775 
& 83\,298 \\

\;\;\;\; {Train} 
& 86\,677 
& 66\,264 
& 78\,089 \\

\;\;\;\; {Test} 
& 10\,807 
& 8\,511 
& 5\,209 \\

\;\;\;\; Min/Avg/Max per user 
& 4/9.01/173
& 4/8.74/193
& 5/15.99/361 \\

\midrule
\textbf{Statements} \\
\;\; Unique 
& 59\,803 
& 63\,333 
& 81\,516 \\

\;\; Min/Avg/Max per inter. 
& 1/2.26/17 
& 1/2.10/22 
& 1/1.98/16 \\

\;\; Min/Avg/Max per user 
& 3/19.92/415 
& 1/17.98/358 
& 5/31.42/823 \\

\;\; Min/Avg/Max per item 
& 4/31.65/592 
& 3/26.20/333 
& 5/41.92/580 \\

\;\; Triplets 
& 220\,262 
& 156\,721 
& 164\,931 \\
\bottomrule
\end{tabular}
}
\vspace{-4mm}
\end{table}
\begin{table}[t]
\caption{Test-user populations by available history length.
History length counts past training and validation
interactions, before token-budget truncation.}
\label{tab:history-populations}
\vspace{-3mm}
\begin{center}
\small
\setlength{\tabcolsep}{6pt}
\begin{tabular}{lrrr}
\toprule
\textbf{Past interactions} & \textbf{Beauty}
& \textbf{Toys} & \textbf{Yelp} \\
\midrule
4--7      & 7\,702 & 6\,178 & 2\,561 \\
8--15     & 2\,186 & 1\,651 & 1\,417 \\
16--31    & 698    & 533    & 666 \\
$\geq$32  & 221    & 149    & 565 \\
\midrule
\textbf{Total} & 10\,807 & 8\,511 & 5\,209 \\
\bottomrule
\end{tabular}
\end{center}
\vspace{-4mm}
\end{table}

\section{Baselines}
\label{app:baselines}
We compare \textsc{LMStaR} against the following baselines, grouped by their underlying approach:

\begin{itemize}[leftmargin=*,nosep]
    \item \textbf{(1) Popularity-based methods.}
    \textsc{ItemPop}, \textsc{GlobalPop}, and \textsc{UserPop}~\cite{kabongo2026rank} rank statements by their frequency in item histories, the training corpus, and user histories, respectively.

    \item \textbf{(2) Neighborhood-based collaborative filtering.}
    \textsc{RUCF} and \textsc{RICF}~\cite{li2021extra} score statements using Jaccard-based similarities between users and items, respectively.

    \item \textbf{(3a) Collaborative factorization.}
    \textsc{CD} and \textsc{PITF}~\cite{li2021extra} model ternary and pairwise latent interactions, respectively. \textsc{BPER}~\cite{li2023relationship} combines user--statement and item--statement scores, while \textsc{BPER-Large}~\cite{xu2022learning} increases its latent dimension to 768. These models are trained with pairwise BPR objectives~\cite{rendle2012bpr}.

    \item \textbf{(3b) Semantically enhanced factorization.}
    \textsc{BPER+}~\cite{li2023relationship} integrates BERT embeddings through element-wise multiplication, whereas \textsc{SE-BPER} and \textsc{SE-BPER+}~\cite{xu2022learning} initialize latent factors from pretrained and task-adapted statement representations, respectively. All retain BPR-based training.

    \item \textbf{(4) Contrastive collaborative learning.}
    \textsc{CLER}~\cite{escarda2026contrastive} learns user, item, and statement ID embeddings using a contrastive NT-BXent objective.

    \item \textbf{(5) Zero-shot neural retrieval.}
    \textsc{Qwen3-Embedding-8B}~\cite{zhang2025qwen3}, a pretrained instruction-aware retrieval model, is evaluated without task-specific fine-tuning.
\end{itemize}

\section{Evaluation Metrics}
\label{app:metrics}
For each user--item interaction $(u,i)$, let $\pi_{ui}(j)$ denote the statement ranked at position $j$ among the candidates in $\mathcal{S}_i$. We define binary relevance as $\mathrm{rel}_{ui}(j) = \mathbb{I}[\pi_{ui}(j) \in \mathcal{S}_{ui}]$, where $\mathbb{I}[\cdot]$ is the indicator function. Ranking quality is measured using Precision@$k$, Recall@$k$, and NDCG@$k$~\cite{jarvelin2002cumulated}, defined as:
\begin{align}
    \mathrm{P}@k(u,i)
    &= \frac{1}{k}\sum_{j=1}^{k}\mathrm{rel}_{ui}(j), \\
    \mathrm{R}@k(u,i)
    &= \frac{1}{|\mathcal{S}_{ui}|}
    \sum_{j=1}^{k}\mathrm{rel}_{ui}(j), \\
    \mathrm{NDCG}@k(u,i)
    &= \frac{1}{\mathrm{IDCG}@k(u,i)}
    \sum_{j=1}^{k}
    \frac{\mathrm{rel}_{ui}(j)}{\log_2(j+1)},
\end{align}
where the ideal discounted cumulative gain is:
\begin{align}
    \mathrm{IDCG}@k(u,i)
    = \sum_{j=1}^{\min(k,|\mathcal{S}_{ui}|)}
    \frac{1}{\log_2(j+1)}.
\end{align}
Dataset-level scores are obtained by averaging the interaction-level metrics over the test set.

\section{Implementation Details}
\label{app:implementation}

\subsection{LMStaR Training Configuration}
\label{app:lmstar-implementation}

Unless otherwise specified, we use the same training
configuration across backbones and datasets, as summarized
in Table~\ref{tab:lmstar-hparams}.
The query and statement encoders are initialized as two
independent copies of the same pretrained backbone and
adapted using LoRA on the query, key, value, and output
attention projections.
We extract the last non-padding token representation and
apply $\ell_2$ normalization to both embeddings.
The query includes the task instruction, target-item
metadata, and the user's interaction history, ordered from
most recent to oldest and including all available statements
for each interaction. We retain the longest history prefix
fitting the query token budget, including the target-item
metadata. Statement inputs are limited to 256 tokens.
Training uses a multi-positive InfoNCE objective with
32 uniformly sampled item-level negatives per interaction.
Checkpoints are evaluated every 50 optimization steps and
selected according to validation NDCG@5, with early stopping
after five evaluations without improvement.

\subsection{Backbone-Specific Settings}
\label{app:lmstar-backbones}

We evaluate five pretrained models:
EmbeddingGemma-300M~\cite{vera2025embeddinggemma}, Qwen3-0.6B~\cite{yang2025qwen3}, and Qwen3-Embedding (0.6B, 4B, 8B)~\cite{zhang2025qwen3}.
In particular, EmbeddingGemma-300M uses bidirectional attention, its
native query and statement prefixes
(\texttt{task: search result | query:} and
\texttt{title: none | text:}), and a 2,048-token query
limit corresponding to its maximum context length.
The Qwen3-based models use causal attention and a
4,096-token query limit.
For the 4B and 8B backbones, the physical batch size is
reduced from 128 to 64 and gradient accumulation is
increased from two to four steps, preserving an effective
batch of 256 interactions.
All experiments use NVIDIA H200 GPUs (141 GB each),
with two GPUs for the 4B and 8B backbones and one GPU
for the smaller models.
For the full fine-tuning comparison in
Table~\ref{tab:lmstar-adaptation}, we update all parameters
of both encoders, rather than applying LoRA, for each of
the two evaluated backbones: EmbeddingGemma-300M and
Qwen3-Embedding-0.6B.

\subsection{Baselines}
\label{app:implementation-baselines}

We follow the original baseline implementations, preserving
their architectures, objectives, and training procedures
wherever possible. The main adaptation concerns negative
sampling, which we align with our item-level ranking setting:
for methods requiring negative statements, these are sampled
exclusively from the target item's statement pool, excluding
positives associated with the corresponding user--item pair.
Only statements observed in the training split are eligible
as training negatives. At evaluation time, all methods rank
the same complete set of candidate statements associated
with the target item.
Each baseline is evaluated on every dataset using three
random seeds (40, 41, and 42), and we report the mean
performance across these runs. Statistical baselines are
executed on CPUs, purely collaborative models (including
\textsc{CLER}) on a single NVIDIA RTX A6000 GPU, and
semantics-enhanced models on a single NVIDIA H200 GPU.
Detailed implementation settings and hyperparameter
configurations are provided in the accompanying code.

\begin{table}[t]
\caption{\textsc{LMStaR} training hyperparameters.}
\label{tab:lmstar-hparams}
\vspace{-4mm}
\begin{center}
\small
\setlength{\tabcolsep}{5pt}
\begin{tabular}{ll}
\toprule
\textbf{Hyperparameter} & \textbf{Value} \\
\midrule
LoRA rank / $\alpha$ / dropout & 16 / 32 / 0 \\
LoRA target modules & Attention $q,k,v,o$ \\
Temperature $\tau$ & 0.1 \\
Negatives per interaction & 32 (item-level) \\
Effective batch & 256 \\
Optimizer / learning rate & AdamW / $10^{-4}$ \\
Weight decay & 0.01 \\
LR schedule / warmup & Cosine / 5\% \\
Gradient clipping (max norm) & 1.0 \\
Maximum epochs & 10 \\
Validation interval / patience & 50 steps / 5 evaluations \\
Checkpoint selection & Validation NDCG@5 \\
Numerical precision & bf16 \\
Random seed & 40 \\
\bottomrule
\end{tabular}
\end{center}
\vspace{-4mm}
\end{table}

\section{Extended Related Work}

Our work lies at the intersection of text-based explainable recommendation and neural retrieval. We review existing approaches to generating and ranking explanations and advances in dense representation learning.

\subsubsection*{Text-based Explainable Recommendation}
Text-based explainable recommendation has evolved from template-based methods~\cite{li2020generate} to neural generation using RNNs~\cite{dong2017learning,li2017neural}, Transformers~\cite{li2021personalized}, and more recently pretrained language models and LLMs~\cite{li2023personalized,ma2024xrec,li2025g}. Despite advances in fluency, evaluation remains challenging: lexical metrics reward surface overlap~\cite{papineni2002bleu,lin2004rouge}, semantic metrics do not ensure factual consistency~\cite{zhang2019bertscore,sellam2020bleurt,kabongo2025factual}, and LLM-based judges can be biased and difficult to reproduce~\cite{fu2023gptscore,sheng2025analyzing}. Moreover, paragraph-level evaluation obscures the relevance of individual explanatory factors. 
To address these limitations, prior work~\cite{li2021extra,kabongo2026rank} shifts from free-form generation to ranking short explanatory passages, called \emph{statements}, selecting the top-$k$ candidates as explanations. This enables standardized evaluation using established ranking metrics~\cite{jarvelin2002cumulated}. Meaningful evaluation further requires statements to be explanatory, atomic, and semantically unique~\cite{kabongo2026rank}.

\subsubsection*{Statement Ranking Methods}
Existing approaches range from statistical baselines to collaborative and semantically enhanced models. Popularity baselines~\cite{kabongo2026rank} rank statements by their frequency in user histories (\textsc{UserPop}), item histories (\textsc{ItemPop}), or the training corpus (\textsc{GlobalPop}). Collaborative approaches exploit user--item--statement interactions: \textsc{RUCF} and \textsc{RICF} aggregate evidence from Jaccard-based neighborhoods, while \textsc{CD} and \textsc{PITF} rely on ternary and pairwise latent factorization, respectively~\cite{li2021extra}. \textsc{BPER}~\cite{li2023relationship} combines separately learned user--statement and item--statement scores.
To incorporate textual semantics, \textsc{BPER+}~\cite{li2023relationship} combines latent factors with BERT~\cite{devlin2019bert} embeddings, whereas \textsc{SE-BPER} and \textsc{SE-BPER+} initialize latent factors from contextualized statement representations, with the latter introducing task-specific fine-tuning~\cite{xu2022learning}. While these models rely on Bayesian Personalized Ranking (BPR) objective~\cite{rendle2012bpr}, \textsc{CLER}~\cite{escarda2026contrastive} adopts a contrastive NT-BXent loss while retaining ID-based representations.

Explanation ranking distinguishes \emph{global-level} ranking over all statements from \emph{item-level} ranking restricted to the target item's statements~\cite{kabongo2026rank, li2021extra}. While the former resembles large-scale retrieval, the latter more directly assesses personalization. Results on the \textsc{StaR} benchmark~\cite{kabongo2026rank} show that popularity baselines remain competitive globally, while \textsc{UserPop} matches or outperforms several state-of-the-art models at the item level, motivating our focus on this setting.

\subsubsection*{Statement Ranking Datasets}
Real-world recommendation datasets contain free-form reviews rather than structured explanatory statements. \textsc{EXTRA}~\cite{li2021extra} constructs explanation-ranking datasets by extracting review sentences and grouping near-duplicates through lexical $n$-gram similarity and locality-sensitive hashing~\cite{anand2011mining}. However, lexical matching may retain non-explanatory content and overlook semantic paraphrases. \textsc{StaR}~\cite{kabongo2026rank} addresses these limitations through LLM-based statement extraction and a three-stage semantic clustering pipeline combining approximate nearest-neighbor retrieval, reranker-based filtering, and graph-based refinement. We build upon this pipeline with two refinements: an extraction prompt emphasizing the user--item match and stricter reranker-based verification of cluster merges. These changes aim to improve explanatory relevance and semantic consistency.

\subsubsection*{Neural Retrieval Models}
Neural retrieval represents queries and candidates in a shared embedding space, enabling efficient similarity-based ranking with precomputable candidate representations. Early dual-encoder approaches, including {DSSM}~\cite{huang2013learning}, {Sentence-BERT}~\cite{reimers2019sentence}, and {DPR}~\cite{karpukhin2020dense}, established this paradigm. Subsequent work improved dense retrieval through contrastive learning and hard negative mining~\cite{xiong2020approximate}, large-scale weak supervision~\cite{wang2022text}, and instruction-conditioned representations~\cite{su2023one}. More recently, {RepLLaMA}~\cite{ma2024fine}, {LLM2Vec}~\cite{behnamghader2024llm2vec}, and {Qwen3-Embedding}~\cite{zhang2025qwen3} have extended dense retrieval to LLM-based representations.
In recommendation, neural candidate retrieval has been widely adopted~\cite{covington2016deep}, while {UniSRec}~\cite{hou2022towards} and {RecFormer}~\cite{li2023text} leverage textual item information and interaction histories to model user preferences. Recent LLM-based two-tower architectures further extend this paradigm to personalized item retrieval~\cite{xu2026case}. Building on these advances, \textsc{LMStaR} formulates personalized explanation ranking as dense retrieval, matching textual user--item contexts with candidate explanatory statements.

\end{document}